\documentclass[11pt,journal,twoside]{IEEEtran}

\usepackage{cite}
\usepackage{theorem,bm,amssymb,amsmath}
\usepackage{paralist}
\usepackage{tikz}
\newcommand*\circled[1]{\tikz[baseline=(char.base)]{
            \node[shape=circle,draw,inner sep=2pt] (char)
                 {$\scriptstyle #1$};}}
\def\graphedge{\mskip-1.5mu\mathchar"0200\mskip-3mu\mathchar"0200\mskip-1.5mu}
\def\graphleftedge{\mskip-1.25mu\mathchar"0220\mskip-3mu\mathchar"0200\mskip-1.5mu}
\def\graphrightedge{\mskip-1.5mu\mathchar"0200\mskip-3mu\mathchar"0221\mskip-1.25mu}

\usepackage{color}
\usepackage{graphicx}
\RequirePackage[pdftex,pdfpagemode=none, pdftoolbar=true,
                pdffitwindow=true,pdfcenterwindow=true]{hyperref}

\definecolor{darkgreen}{rgb}{0,0.5,0}
\definecolor{darkyellow}{rgb}{0.5,0.5,0}
\definecolor{darkred}{rgb}{0.6667,0,0}
\definecolor{light-gray}{gray}{0.5}

\makeatletter

\let\IfDraftVersion\if@draftclsmode

\def\threeauthors#1#2#3#4#5#6{\gdef\@address{}
   \gdef\@name{\begin{tabular}{@{}c@{}}
        {\em #1}\\\noalign{\vskip 6pt plus 3pt minus 3pt}
        #2\relax
   \end{tabular}\hskip .5in plus.5in minus.125in\begin{tabular}{@{}c@{}}
        {\em #3}\\\noalign{\vskip 9pt plus 3pt minus 3pt}
        #4\relax
   \end{tabular}\hskip .5in plus.5in minus.125in\begin{tabular}{@{}c@{}}
        {\em #5}\\\noalign{\vskip 6pt plus 3pt minus 3pt}
        #6\relax
\end{tabular}}}

\def\Rbbb{{\mathbb{R}}}

\def\Psib{{\bm{\Psi}}}

\def\Thetab{{\bm{\Theta}}}

\def\alphab{{\bm{\alpha}}}

\def\phib{{\bm{\phi}}}
\def\pib{{\bm{\pi}}}

\def\varthetab{{\bm{\vartheta}}}
\def\xib{{\bm{\xi}}}

\def\Ab{{\mathbf{A}}}
\def\Bb{{\mathbf{B}}}

\def\Db{{\mathbf{D}}}

\def\Ib{{\mathbf{I}}}

\def\Kb{{\mathbf{K}}}
\def\Gb{{\mathbf{G}}}
\def\Lb{{\mathbf{L}}}
\def\Mb{{\mathbf{M}}}

\def\Qb{{\mathbf{Q}}}

\def\Sb{{\mathbf{S}}}

\def\Xb{{\mathbf{X}}}

\def\db{{\mathbf{d}}}

\def\lb{{\mathbf{l}}}
\def\nb{{\mathbf{n}}}

\def\sb{{\mathbf{s}}}

\def\vb{{\mathbf{v}}}

\def\xb{{\mathbf{x}}}

\def\zb{{\mathbf{z}}}

\def\zerob{{\bm{0}}}
\def\oneb{{\bm{1}}}

\def\LPolishb{\mbox{\bf\L}}

\DeclareMathOperator{\diag}{diag}
\DeclareMathOperator{\Diag}{Diag}

\DeclareMathOperator{\PD}{PD}
\DeclareMathOperator{\PFA}{PFA}

\def\T{{\scriptscriptstyle\rm T}}   
\let\humlaut=\H
\def\H{\ifmmode{\scriptscriptstyle\rm H}\else\humlaut\fi}   

\def\by{\ifmmode $\hbox{-by-}$\else \leavevmode\hbox{-by-}\fi}
\def\sqrtm1{{\sqrt{\!-1}}}
\let\(=\langle
\let\)=\rangle

\def\detthresh{\mathrel{\mathop{\gtrless}\limits_{{\rm H}_0}^{{\rm H}_1}}}
\def\detthreshv#1{\mathrel{\mathop{\gtrless}\limits_{{\rm H}_0(#1)}^{{\rm H}_1(#1)}}}

\def\hgfarg#1{\left(\null\vcenter{\normalbaselines\m@th
    \ialign{&$\displaystyle##$\hfil\crcr
      \mathstrut\crcr\noalign{\kern-\baselineskip}
      #1\crcr\mathstrut\crcr\noalign{\kern-\baselineskip}}}\right)}
\def\Heqalign#1{\null\,\vcenter{\openup\jot\m@th
  \ialign{\strut\hfil$##$:\quad&\hfil$\displaystyle{##}$&$\displaystyle
      {{}##}$\hfil&\qquad##\hfil\crcr#1\crcr}}\,}
\def\eqalignno#1{\displ@y \tabskip\@centering
  \halign to\displaywidth{\hfil$\@lign\displaystyle{##}$\tabskip\z@skip
    &$\@lign\displaystyle{{}##}$\hfil\tabskip\@centering
    &\llap{$\@lign##$}\tabskip\z@skip\crcr
    #1\crcr}}
\def\dbleqalignno#1{\displ@y \tabskip\@centering
  \halign to\displaywidth{\hfil$\@lign\displaystyle{##}$\tabskip\z@skip
    &$\@lign\displaystyle{{}##}$\hfil
    &$\@lign\displaystyle{{}##}$\hfil\tabskip\@centering
    &\llap{$\@lign##$}\tabskip\z@skip\crcr
    #1\crcr}}
\def\properties#1{\openup\jot\mathsurround=0pt\tabskip=0pt
  \everycr{\noalign{\nobreak}}
  \halign to\hsize{\strut\indent\indent\llap{##}\enskip\tabskip=0pt plus1fil&%
    \hfil$\displaystyle{##}$\tabskip=0pt&$\displaystyle{{}##}$\hfil
      \tabskip=0pt plus1fil&\qquad##\hfil\tabskip=0pt
      \crcr#1\crcr}}

\theoremstyle{plain}

\newif\ifmathtomb \mathtombfalse

\def\tombstone{\unskip\penalty50   
  \hskip 0pt plus-1fill \null\nobreak\hskip 0pt plus1fill
  \enskip \vrule width.3333em height.7em depth.2em
  \ifmmode \global\mathtombtrue \else \global\mathtombfalse \fi}

  {\futurelet\next\hpr@oftext}
  {\ifmathtomb \else \tombstone \fi \widowpenalty=10000  
   \par \ifmathtomb \else \addvspace{\medskipamount}\fi \global\mathtombfalse}
\def\hpr@oftext{\ifx\next[\let\temp\ohpr@@ftext\else\let\temp\hpr@@ftext\fi\temp}
\def\hpr@@ftext{\beginhpr@@f{Proof}}
\def\ohpr@@ftext[#1]{\beginhpr@@f{#1}}
\def\beginhpr@@f#1{\par \addvspace{\bigskipamount}%
  \noindent{\bf #1:\enspace}\ignorespaces }

\setbox\strutbox=\hbox{\vrule height9pt depth4pt width\z@}%
\def\big#1{{\hbox{$\left#1\vbox to9.5\p@{}\right.\n@space$}}}%
\def\Big#1{{\hbox{$\left#1\vbox to12.5\p@{}\right.\n@space$}}}%
\def\bigg#1{{\hbox{$\left#1\vbox to16\p@{}\right.\n@space$}}}%
\def\Bigg#1{{\hbox{$\left#1\vbox to19\p@{}\right.\n@space$}}}%

\makeatother

\begin{document}


\let\originalnormalsize=\normalsize
\title{Network Detection Theory and Performance}

\author{Steven T.~Smith*,~\IEEEmembership{Senior~Member, IEEE},
  Kenneth~D. Senne*,~\IEEEmembership{Life~Fellow, IEEE},\\
  Scott Philips*, Edward K. Kao*\dag, and Garrett Bernstein*
\thanks{Manuscript received 2013.}
\thanks{*MIT Lincoln Laboratory, Lexington, MA 02420;
    \{\thinspace\href{mailto:stsmith@ll.mit.edu}{stsmith},
    \href{mailto:senne@ll.mit.edu}{senne},
    \href{mailto:garrett.bernstein@ll.mit.edu}{garrett.bernstein}\thinspace\}@\penalty50ll.mit.edu,\hskip0.5em
    \href{mailto:edwardkao@fas.harvard.edu}{edwardkao@fas.harvard.edu}}
\thanks{\dag Department of Statistics, Harvard University; Cambridge MA USA 02138}
}

\markboth{Smith et~al.: Network Detection}{Smith et~al.: Network
  Detection}

\maketitle

\begin{abstract}
Network detection is an important capability in many areas of applied
research in which data can be represented as a graph of entities and
relationships.  Oftentimes the object of interest is a relatively
small subgraph in an enormous, potentially uninteresting background.
This aspect characterizes network detection as a ``big data'' problem.
Graph partitioning and network discovery have been major research
areas over the last ten years, driven by interest in internet search,
cyber security, social networks, and criminal or terrorist activities.
The specific problem of network discovery is addressed as a special
case of graph partitioning in which membership in a small subgraph of
interest must be determined.  Algebraic graph theory is used as the
basis to analyze and compare different network detection methods.  A
new Bayesian network detection framework is introduced that partitions
the graph based on prior information and direct observations. The new
approach, called space-time threat propagation, is proved to maximize
the probability of detection and is therefore optimum in the
Neyman-Pearson sense.  This optimality criterion is compared to
spectral community detection approaches which divide the global graph
into subsets or communities with optimal connectivity properties. We
also explore a new generative stochastic model for covert networks and
analyze using receiver operating characteristics the detection
performance of both classes of optimal detection
techniques.\let\thefootnote\relax\footnotetext{*This work is sponsored
  by the Assistant Secretary of Defense for Research \& Engineering
  under Air Force Contract FA8721-05-C-0002.  Opinions,
  interpretations, conclusions and recommendations are those of the
  author and are not necessarily endorsed by the United States
  Government.}\end{abstract}


\overfullrule 5pt
\sloppy

\section{Introduction}
\label{sec:intro}  

\begin{figure*}[t]
\medskip
\normalsize
\centerline{\includegraphics[width=0.9167\linewidth]{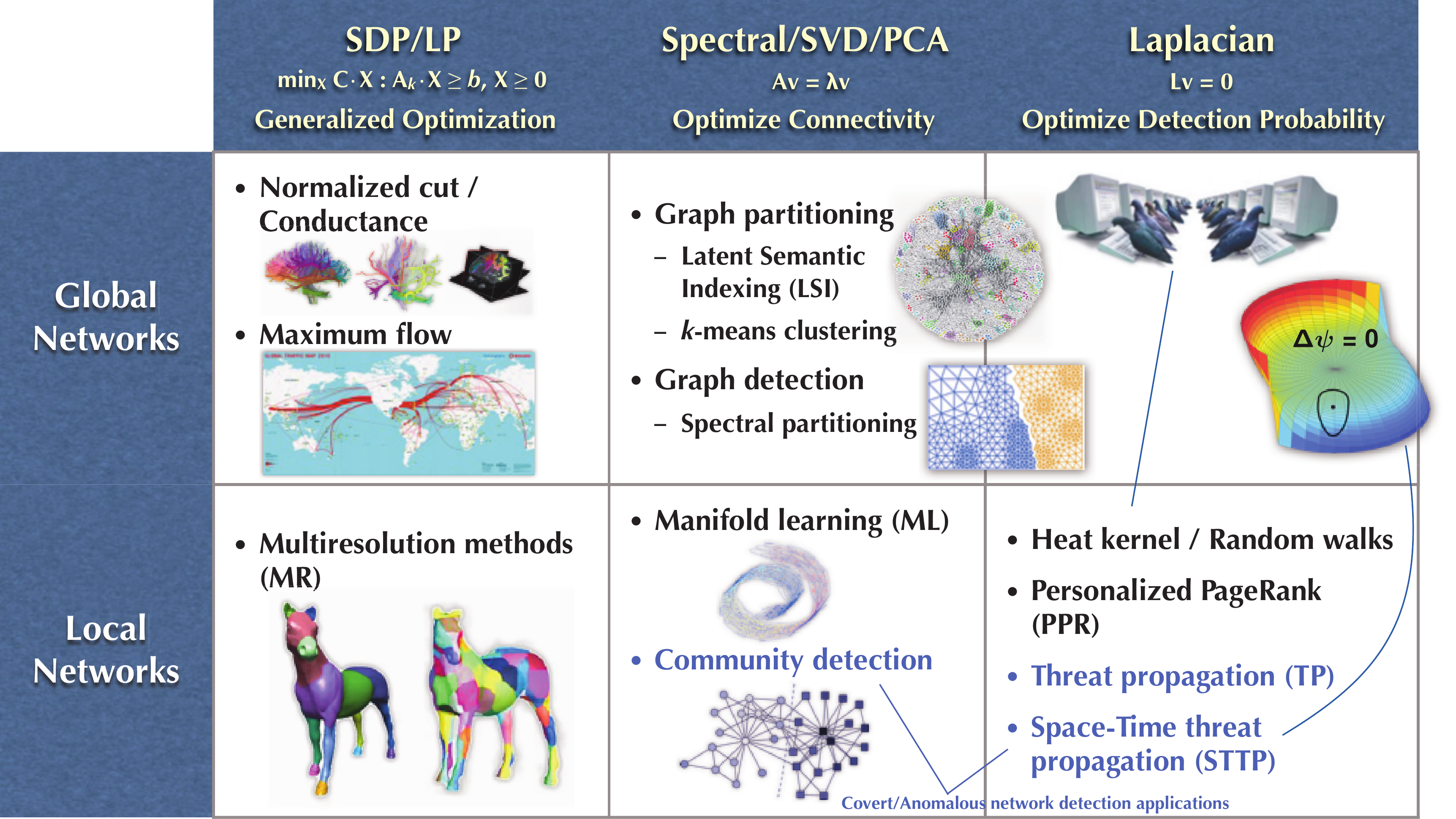}}
\caption{Network detection algorithm taxonomy. This paper focuses on
  local spectral and harmonic methods for network detection. (Images
  used from various sources; clockwise from upper left
  [\citen{Brun2004},\thinspace \citen{TeleGeography2010},\thinspace
    \citen{Guruharsha2011},\thinspace \citen{Newman2006},\thinspace
    \citen{Google2002},\thinspace \citen{Shamir2008},\thinspace
    \citen{Yang2007}]).\label{fig:netdetalg}}
\end{figure*}

Network detection is a special class of the more general graph
partitioning (GP) problem in which the binary decision of membership
or non-membership for each graph vertex must be determined.  This
detection problem and more generally GP are of fundamental and
practical importance in graph theory and its applications
(Figure~\ref{fig:netdetalg}).  The detected subgraph comprises all
vertices declared to be members. The very definition of membership
will lead to specific network detection algorithms.

Graph partitioning is an NP-hard problem; however, semidefinite
programming (SDP) relaxation applies to many cases, offering both
practical and oftentimes theoretically attractive approximation to
GP.\cite{Wolkowicz1999,Leskovec2010} In general, practical GP
approaches exploit a variety of global and local connectivity
properties to divide a graph into many subgraphs.  Decreasing
algorithmic complexity is achieved in certain domains that may be cast
as quadratic optimization problems (yielding eigenvalue- or
spectral-based methods), or simple sets of linear equations.  One
important network detection approach, called community detection,
divides the global graph into subsets or communities based on
optimizing a specific connectivity measure that is chosen depending
upon the application.  This paper presents a new Bayesian network
detection approach called space-time threat
propagation~\cite{Philips2012,Smith2012} that is shown to optimize the
probability of network detection in a Neyman-Pearson sense given prior
information and\slash or direct observations, i.e.\ detection
probability is maximized given a fixed false alarm a.k.a.\ false
positive probability. This is an important property because it
provides a practical optimum algorithm in many settings (satisfying a
set of assumptions detailed later in the paper), and it provides a
performance bound on detection performance. Remarkably, the two
apparently different optimal network detection approaches are related
to each other using insights from algebraic graph theory.  Converse
to other research on network detection, rather than using the network
to detect signals,~\cite{Chamberland2003,Alanyali2004,Kar2008} the
signal of interest in this paper {\em is\/} the signal to be
detected. In this sense the paper is also related to work on so-called
manifold learning methods,~\cite{Belkin2003,Costa2004,Carter2010}
although the network to be detected is a subgraph of an existing
network, and therefore the methods described here belong to a class of
network anomoly detection~\cite{Carter2010} as well as
maximimum-likelihood methods for network detection.~\cite{Ferry2009}
Both spectral-based and Neyman-Pearson network detection methods are
described and analyzed below in Sections \ref{sec:optnetdet}
and~\ref{sec:netdetperf}. Furthermore, network detection performance
is assessed using a new stochastic blockmodel~\cite{Airoldi2008} for
small, dynamic foreground networks embedded within a large background.

\subsection{Covert Networks}
\label{sec:covert}

Detection of network communities is most likely to be effective if the
communities exhibit high levels of connection activity. However, the
covert networks of interest to many applications are unlikely to
cooperate with this optimistic assumption. Indeed, a ``fully connected
network \dots\ is an unlikely description of the enemy insurgent order
of battle.''\cite{USArmy2006} A clandestine or covert community is
more likely to appear cellular and distributed.\cite{Carley2004}
Communities of this type can be represented with ``small world''
models.\cite{Sageman2004} The covert networks of interest in this
paper exist to accomplish nefarious, illegal, or terrorism goals,
while ``hiding in plain sight.''\cite{Jonas2006,Xu2008} Covert networks
necessarily adopt operational procedures to remain hidden and robustly
adapt to losses of parts of the network. For example, during the
Algerian Revolution the FLN's Autonomous Zone of Algiers (Z.A.A.)
military command was ``carefully kept apart from other elements of the
organization, the network was broken down into a number of quite
distinct and compartmented branches, in communication only with the
network chief,'' allowing Z.A.A. leader Yassef Saadi to command
``within 200 yards from the office of the [French] army commandant
\dots\ and remain there several months[.]''\cite{Trinquier2006} Krebs'
reconstruction of the 9/11 terrorist network details the strategy for
keeping cell members distant from each other and from other cells and
notes bin~Laden's description of this organization: ``those \dots\ who
were trained to fly didn't know the others.  One group of people did
not know the other group.''\cite{Krebs2002} A covert network does not
have to be human to be nefarious; the widespread Flashback malware
attack on Apple's OS~X computers employed switched load balancing
between servers to avoid detection,\cite{DrWeb2012} mirroring the
Z.A.A.'s ``tree'' structure for robust covert network organization.

In order to accomplish its goals the covert network must judiciously
use ``transitory shortcuts.''\cite{Watts1999} For example, in the 9/11
terrorism operation, after coordination meetings connected distant
parts of the network, the ``cross-ties went dormant.''\cite{Krebs2002}
It is during these occasional bursts of connection activity that a covert
community may be most vulnerable to detection.\cite{USArmy2006}

Network detection is predicated on the existence of observations of
network relationships. In this paper the focus will be on observations
of network activities using Intelligence, Surveillance, and
Reconnaissance (ISR) sensors, such as Wide-Area Motion Imagery
(WAMI)\null.  Covert networks engaged in terrorist attacks with
Improvised Explosive Devices (IEDs) comprise loosely connected cells
with various functions, such as finance, planning, operations,
logistics, security, and propaganda.

In this paper a new model of covert threat for detection analysis that
accounts for the realities of dynamic foreground networks in large
backgrounds is a specially adapted version of a mixed membership
stochastic blockmodel.\cite{Airoldi2008} The terrorist cells of
interest are embedded into a background consisting of many ``neutral''
communities, that represent business, homes, industry, religion,
sports, etc. Because in real life people wear different ``hats''
depending upon on the communities with which they interact, their
proportions of membership in multiple communities (lifestyles) can be
adjusted to control the occasional coordination between the foreground
and background networks. The new generative blockmodel approach
introduced in Section~\ref{sec:covnetsb} leads to a analytically
tractable tool with sufficient parameters to exhibit realistic
coordinated activity levels and interactions.

\subsection{Observability and Detectability}
\label{sec:obsdet}

The connections (edges) between nodes of a network are observable
only when they are active.  This implies that there are two basic
strategies for detecting a covert threat: (1)~subject-based Bayesian
models that correlate a~priori information or observations of the
observed network connections; (2)~pattern-based (predictive) methods
that look for known patterns of organization\slash behavior to infer
nefarious activity.\cite{Jonas2006,Perry2008} Subject-based methods
follow established principles of police investigations to accrue
evidence based upon observed connections and historical data.  The
dependency of predictive methods on known patterns, however, makes
them difficult to apply to rare and widely different covert threats:
``there are no meaningful patterns that show what behavior indicates
planning or preparation for terrorism.''\cite{Jonas2006} The
real-world consequences of applying an inappropriate model to detect a
threat may include an unacceptable number of false positives and an
erosion of individual privacy rights and civil
liberties.\cite{Jonas2006,Perry2008}

As described above, the subject of community detection in graphs has
experienced extensive research during the last ten
years.\cite{Newman2006,Jackson2008,Koller2009,Fortunato2010,Newman2010}
Nevertheless, there are few closed-form results that quantify the
limits of detectability of specific types of networks in
representative backgrounds.  Fully connected networks (cliques) have
received special attention: there is a recent result which confirms in
closed form using random matrix theory the previously observed phase
transition of detectability for sufficiently small
cliques~\cite{Fortunato2007,Kumpula2007,Nadakuditi2012} or dense
subgraphs~\cite{Arias-Castro2013}.

In this paper we use the proposed generative stochastic threat model
with Monte~Carlo detection performance analysis.  The detection
methodologies under investigation here include the spectral-based and
Neyman-Pearson techniques discussed above in the Introduction.

\section{Algebraic Graph Theory}
\label{sec:agt}

A graph~$G=(V,E)$ is defined by two sets, the vertices $V$ of~$G$, and
the edges $E\subset[V]^2\subset2^V$ of~$G$, in which $[V]^2$ denotes
the set of $2$-element subsets of~$V$.\cite{Diestel2000} For example,
the sets $V=\{\,1,\;2,\;3\,\}$,
$E=\bigl\{\,\{1,\,2\},\;\{2,\,3\}\,\bigr\}$ describe a simple graph
with undirected edges between vertices $1$ and~$2$, and $2$ and~$3$:
$\circled{1}\graphedge\circled{2}\graphedge\circled{3}$. The {\em
  adjacency matrix\/} ${\Ab=\Ab(G)}$ of~$G$ is the $\{0,1\}$-matrix
with ${\Ab_{ij}=1}$ iff $\{\,i,j\,\}\in E$. In the example, $\Ab=
\left(\begin{smallmatrix}
  0&1&0\\ 1&0&1\\ 0&1&0\end{smallmatrix}\right)$. Because simple
  graphs are undirected, their adjacency matrix is necessarily
  symmetric. The {\em degree matrix\/} ${\Db=\Diag(\Ab{\cdot}\oneb)}$
  is the diagonal matrix of the vector of degrees of all vertices,
  where ${\oneb=(1,\ldots,1)^\T}$ is the vector of all ones.

Many important applications involve an orientation between vertices,
defined by an orientation map $\sigma\colon[V]^2\to {V\times V}$ (the
ordered Cartesian product of~$V$ with itself) in which the first and
second coordinates are called the initial and terminal vertices,
respectively. The corresponding {\em directed graph\/} is denoted
$G^\sigma$ or, by abuse of notation, simply $G$. The preceding example
with orientation map ${\sigma(\{1,\,2\})=(2,1)}$,
${\sigma(\{2,\,3\})=(2,3)}$ yields the directed graph
$\circled{1}\graphleftedge\circled{2}\graphrightedge\circled{3}$. The
{\em incidence matrix\/} ${\Bb=\Bb(G^\sigma)}$ of the oriented graph
$G^\sigma$ is the $(0,\pm1)$-matrix of size $\#V\by\#E$ with
${\Bb_{ie}=-1}$ if $i$ is an initial vertex of~$\sigma(e)$, $1$ if $i$
is a terminal vertex of~$\sigma(e)$, and $0$ otherwise. In the
example, $\Bb= \left(\begin{smallmatrix}
  \hphantom{-}1&\hphantom{-}0\\ -1&-1\\ \hphantom{-}0&\hphantom{-}1\end{smallmatrix}\right)$. In
  the study of homology in algebraic topology, the incidence matrix is
  recognized as the boundary operator on graph edges. It encodes
  differences between vertices and plays an important role in the
  analysis of network detection algorithms through the so-called graph
  Laplacian, which appears in three forms. The {\em unnormalized
    Laplacian matrix\/} or {\em Kirchhoff matrix\/} of a graph~$G$ is
  the matrix
  \begin{equation}{\Qb=\Qb(G)}=\Bb\Bb^\T=\Db-\Ab,\label{eq:kirchhoff}\end{equation}
where $\Bb(G^\sigma)$ is the incidence matrix of an oriented
graph~$\Gb^\sigma$ with (arbitrary) orientation~$\sigma$, and $\Ab(G)$
and~$\Db(G)$ are, respectively, the adjacency and degree matrices
of~$G$. In the example, $\Qb= \left(\begin{smallmatrix}
  \hphantom{-}1&-1&\hphantom{-}0\\ -1&\hphantom{-}2&-1\\ \hphantom{-}0&-1&\hphantom{-}1\end{smallmatrix}\right)$. The
  (normalized) {\em Laplacian matrix\/}
\begin{equation}\Lb
    =\Db^{-1/2}\Qb\Db^{-1/2}=\Ib-\Db^{-1/2}\Ab\Db^{-1/2}\label{eq:laplacian}\end{equation}
is a matrix congruence of the Kirchhoff matrix~$\Qb$ scaled by the
square-root of the degree matrix~$\Db^{1/2}$. The {\em generalized\/}
or {\em asymmetric Laplacian matrix\/}
\begin{equation} \LPolishb =\Db^{-1/2}\Lb\Db^{1/2} =\Db^{-1}\Qb
  =\Ib-\Db^{-1}\Ab \label{eq:genlaplacian}\end{equation} is a
similarity transformation of the Laplacian matrix. In the example,
$\Lb= \left(\begin{smallmatrix}
  1&-2^{-1/2}&0\\ -2^{-1/2}&1&-2^{-1/2}\\ 0&-2^{-1/2}&1\end{smallmatrix}\right)$
  and $\LPolishb= \left(\begin{smallmatrix}
    1&-1&0\\ -2^{-1}&\hphantom{-}1&-2^{-1}\\ 0&-1&1\end{smallmatrix}\right)$. The
latter example is immediately recognized as a discretization of the
second derivative $-d^2\!/\!dx^2$, i.e.\ the negative of the
\hbox{$1$-d} Laplacian operator $\Delta =\partial^2\!/\partial x^2
+\partial^2\!/\partial y^2+\cdots$ that appears in numerous physical
applications. (This sign is the convention used in graph theory.) The
asymmetric Laplacian ${\LPolishb=\Ib-\Db^{-1}\Ab}$ plays an important
role in mean-value theorems involving solutions to Laplace's equation
${\LPolishb\vb=\zerob}$, which will be seen to be the motivating
equation behind several network detection algorithms.

The connection between the incidence and Laplacian matrices and
physical applications is made through Green's first identity, which
equates the continuous Laplacian operator $\Delta$ in terms of the
vector gradient $\nabla=(\partial\!/\!\partial x,\partial\!/\!\partial
y,\ldots)^\T$ and motivates the definition ${\Qb=\Bb\Bb^\T}$ of the
graph Laplacian. Given two arbitrary ``test'' functions $f(\xb)$
and~$g(\xb)$ on a bounded domain $\Omega\subset\Rbbb^n$ with boundary
$\partial\Omega$ and inner product~${\langle\,{,}\,\rangle}$, Green's
first identity asserts, \begin{equation} \int_\Omega g\Delta f\,dV
  =-\int_\Omega\langle\nabla g,\nabla\!f\rangle\,dV
  +\int_{\partial\Omega}g\langle\nabla\!f,\nb\rangle\,dS, \label{eq:Greens1st}
  \end{equation} where $dV$ and $\nb\,dS$ are the volume and directed surface
differentials---this formula generalizes immediately to Riemannian
manifolds.  Applying the finite element method to this continuous
equation yields a graph arising from, say, Delaunay triangulation and
a matrix equation involving the graph Laplacian matrix~$\Lb$ [from
  $g\Delta f$ in Eq.~(\ref{eq:Greens1st})] and the normalized outer
product $\Db^{-1/2}\Bb\Bb^\T\Db^{-1/2}$ of the incidence matrix [from
  $\langle\nabla g,\nabla\!f\rangle$ in
  Eq.~(\ref{eq:Greens1st})]. This illustrates that the graph Laplacian
is the standard Laplacian of physics and mathematics, a connection
that explains many theoretical and performance advantages of the
normalized Laplacian over the Kirchhoff matrix across
applications.\cite{Chung1994,Shi2000,vonLuxburg2005,Weiss1999,White2005}

The most important property of the Laplacian matrix is that the
constant vector ${\oneb=(1,\ldots,1)^\T}$ is in the kernel of the
Laplacian, \begin{equation}\Qb\oneb=\zerob;\qquad
  \LPolishb\oneb=\zerob, \end{equation} i.e.\ $\oneb$ is an
eigenvector of~$\Qb$ and~$\LPolishb$ whose eigenvalue is zero. This
property is the reason for the mean-value property of harmonic
functions, as well as the fact that the only bounded harmonic
functions on an unbounded domain are necessarily constant, which will
play an important role in optimum network detection.  This is a key
fact because many network detection algorithms involve solutions to
Laplace's equation, however this constant solution does not
distinguish between vertices at all, a deficiency that may be resolved
in a variety of ways, yielding a family of network detection
algorithms. Furthermore, the geometric multiplicity of the zero
eigenvalue equals the number of connected components of the graph,
though because a connected graph is implicit for the subgraph
detection problem, we may assume that the kernel of the graph
Laplacian is simply the one-dimensional subspace
${(\oneb)=\{\,\alpha\oneb:\alpha\in\Rbbb\,\}}$.

\section{Optimum Network Detection}
\label{sec:optnetdet}

Two different optimality criteria are used for the two
different strategies of network detection: various connectivity
metrics are used for predictive methods, and detection performance is
used for subject-based methods.  Detection optimality means, as usual,
optimality in the Neyman-Pearson sense in which the probability of
detection is maximized at a fixed false alarm rate. In the context of
networks, the probability of detection (PD) refers to the fraction of
vertices detected belonging to the threat subgraph, and the
probability of false alarm (PFA) refers to the fraction of non-threat
vertices detected. As in classical detection
theory,\cite{VanTrees1968} the optimal detector is a threshold of the
log-likelihood ratio (LLR), and a new Bayesian framework for network
detection is developed in this section. The distinction between
classical detection theory and network detection theory is not in the
form of the optimal detector---the log-likelihood ratio---but in
distinct mathematical formulations. Whereas linear algebra is the
foundation for classical detection theory, algebraic graph
theory\cite{Godsil2001} is the foundation for network detection. It
follows that understanding the theory, algorithms, and results of
network detection requires an introduction of some basic concepts from
algebraic graph theory, especially the graph Laplacian and spectral
analysis of graphs. Familiarization with these objects provides a
common framework of comparing apparently unrelated network detection
algorithms and provides deep insights into basic problems in network
detection theory.

\subsection{Spectral-Based Community Detection}
\label{sec:sbcd}

Efficient graph partitioning algorithms and analysis appeared in the
1970s with Donath and Hoffman's eigenvalue-based bounds for graph
partitioning~\cite{Donath1973} and Fiedler's connectivity analysis and
graph partitioning algorithm~\cite{Fiedler1973,Fiedler1975} which
established the connection between a graph's algebraic properties and
the spectrum of its Kirchhoff Laplacian matrix~${\Qb=\Db-\Ab}$
[Eq.~(\ref{eq:kirchhoff})]. The spectral methods in this section solve
the graph partitioning problem by optimizing various subgraph
connectivity properties.

The {\em cut size\/} of a subgraph---the number of edges necessary to
remove to separate the subgraph from the graph---is quantified by the
quadratic form~$\sb^\T\Qb\sb$, where $\sb=(\pm1,\ldots,\pm1)^\T$ is a
$\pm1$-vector who entries are determined by subgraph
membership.~\cite{Pothen1990} Minimizing this quadratic form
over~$\sb$, whose solution is an eigenvalue problem for the graph
Laplacian, provides a network detection algorithm based on the model
of minimal cut size. However, there is a paradox in the application of
spectral methods to network detection: the smallest eigenvalue of the
graph Laplacian ${\lambda_0(\Qb)=0}$ corresponds to the
eigenvector~$\oneb$ constant over~all vertices, which fails to
discriminate between subgraphs. Intuitively this degenerate constant
solution makes sense because the two subgraphs with minimal (zero)
subgraph cut size are the entire graph itself (${\sb\equiv\oneb}$), or
the null graph (${\sb\equiv-\oneb}$). This property manifests itself
in many well-known results from complex analysis, such as the maximum
principle.

Fiedler showed that if rather the eigenvector~$\xib_1$ corresponding
to the second smallest eigenvalue $\lambda_1(\Qb)$ of~$\Qb$ is used
(many authors write ${\lambda_1=0}$ and~$\lambda_2$ rather than the
zero offset indexing ${\lambda_0=0}$ and~$\lambda_1$ used here), then
for every nonpositive constant~${c\le 0}$, the subgraph whose vertices
are defined by the threshold ${\xib_1\ge c}$ is necessarily
connected. This algorithm is called {\em spectral detection}.  Given a
graph~$G$, the number $\lambda_1(\Qb)$ is called the {\em Fiedler
  value\/} of~$G$, and the corresponding eigenvector $\xib_1(\Qb)$ is
called the {\em Fiedler vector}. Completely analogous with comparison
theorems in Riemannian geometry that relate topological properties of
manifolds to algebraic properties of the Laplacian, many graph
topological properties are tied to its Laplacian. For example, the
graph's diameter~$D$ and the minimum degree $d_{\rm min}$ provide
lower and upper bounds for the Fiedler value $\lambda_1(\Qb)$:
$4/(nD)\le \lambda_1(\Qb)\le n/({n-1}){\cdot}d_{\rm
  min}$.\cite{Mohar1991} This inequality explains why the Fiedler
value is also called the {\em algebraic connectivity}: the greater the
Fiedler value, the smaller the graph diameter, implying greater graph
connectivity. If the normalized Laplacian $\Lb$ of
Eq.~(\ref{eq:laplacian}) is used, the corresponding inequality
involving the generalized eigenvalue
${\lambda_1(\Lb)=\lambda_1(\Qb,\Db)}$ involves the graph's
diameter~$D$ and volume~$V$: $1/(DV)\le \lambda_1(\Lb)\le
n/({n-1})$.\cite{Chung1994}

Because in practice spectral detection with its implicit assumption of
minimizing the cut size oftentimes does not detect intuitively
appealing subgraphs, Newman introduced the alternate criterion of
subgraph ``modularity'' for subgraph detection.~\cite{Newman2006}
Rather than minimize the cut size, Newman proposes to maximize the
subgraph connectivity relative to background graph connectivity, which
yields the quadratic maximization problem~$\max_\sb\sb^\T\Mb\sb$,
where ${\Mb=\Ab-V^{-1}\db\db^\T}$ is Newman's {\em modularity matrix},
$\Ab$ is the adjacency matrix, ${(\db)_i=d_i}$ is the degree vector,
and $V=\oneb^\T\db$ is the graph volume.\cite{Newman2006} Newman's
modularity-based graph partitioning algorithm, also called community
detection, involves thresholding the values of the principal
eigenvector of~$\Mb$. Miller
et~al.~\cite{Miller2011,Miller2010a,Miller2010b} also consider
thresholding arbitrary eigenvectors of the modularity matrix, which by
the Courant minimax principle biases the Newman community detection
algorithm to smaller subgraphs, a desirable property for many
applications. They also outline an approach for exploiting
observations within the spectral framework.  \cite{Miller2011}

\subsection{Neyman-Pearson Subgraph Detection}
\label{sec:npnt}

Network detection of a subgraph within a graph ${G=(V,E)}$ of
order~$n$ is treated as $n$ independent binary hypothesis tests to
decide which of the graph's $n$ vertices does not belong (null
hypothesis $H_0$) or belongs (hypothesis $H_1$) to the
network. Maximizing the probability of detection (PD) for a fixed
probability of false alarm (PFA) yields the Neyman-Pearson test
involving the log-likelihood ratio of the competing hypothesis. We
will derive this test in the context of network detection, which both
illustrates the assumptions that ensure detection optimality, as well
as indicates practical methods for computing the log-likelihood ratio
test and achieving an optimal network detection algorithm. It will be
seen that a few basic assumptions yield an optimum test involving the
graph Laplacian, which allows comparison of Neyman-Pearson testing to
several other network detection methods whose algorithms are also
related to the properties of the Laplacian.

Assume that each vertex ${v\in V}$ has an unknown $\{0,1\}$-valued
property $\Theta_v$ which is considered to be ``threat'' or
``non-threat'' at~$v$, and that there exists an observation vector
$\zb\colon \{v_{i_1},\ldots,v_{i_k}\}\subset V\to M\subset\Rbbb^k$
from $k$~vertices to a measurement space~$M$. For example, a direct
observation of threat at vertex~$v$ may be represented by the
observation ${z(v)\equiv1}$. It is assumed that the observation $z(v)$
at~$v$ and the threat $\Theta_v$ at~$v$ are not independent,
i.e.\ $f\bigl(z(v)|\Theta_v\bigr)\ne f\bigl(z(v)\bigr)$, so that there
is positive mutual information between $z(v)$ and~$\Theta_v$. The
probability density $f\bigl(z(v)|\Theta_v\bigr)$ is called the {\em
  observation model}, which in this paper is treated as a simple
$\{0,1\}$-valued model ${\delta\bigl(z(v)-\Theta_v\bigr)}$. Though the
threat network hypotheses are being treated here independently at each
vertex, this framework allows for more sophisticated global models
that include hypotheses over two or more vertices.

An optimum hypothesis test is now derived for the presence of a
network given a set of observations $\zb$. Optimality is defined in
the Neyman-Pearson sense in which the probability of detection is
maximized at a constant false alarm rate (CFAR). As
usual,\cite{VanTrees1968} the derivation of the optimum test involves
the procedure of Lagrange multipliers. For the general problem of
network detection of a subgraph within graph $G$ of order~$n$, the
decision of which of the $2^n$ hypothesis
${\Thetab=(\Theta_{v_1},\ldots,\Theta_{v_n})^\T}$ to choose involves a
$2^n$-ary multiple hypothesis test over the measurement space of the
observation vector~$\zb$, and an optimal test involves partitioning
the measurement space into $2^n$ regions yielding a maximum PD. This
NP-hard general combinatoric problem is clearly computationally and
analytically intractable; however, the general $2^n$-ary multiple
hypothesis test may be greatly simplified by treating it as $n$
independent binary hypothesis tests.

At each vertex ${v\in G}$ and unknown threat $\Theta\colon
V\to\{0,1\}$ across the graph , consider the binary hypothesis test
for the unknown value $\Theta_v$, \begin{equation}\Heqalign{\rm
    H_0(v)& \Theta_v=0&&(vertex belongs to background)\cr \rm H_1(v)&
    \Theta_v=1&&(vertex belongs to
    subgraph).\cr}\label{eq:graphhyp} \end{equation} Given the
observation vector $\zb\colon \{v_{i_1},\ldots,v_{i_k}\}\subset V\to
M\subset\Rbbb^k$ with observation models
$f\bigl(z(v_{i_j})|\Theta_{v_{i_j}}\bigr)$, ${j=1}$, \ldots,~$k$, the
PD and PFA are given by the integrals \begin{align}\hbox{PD}&= \int_R
  f(\zb|{\Theta_v=1})\,d\zb, \label{eq:PD-R}\\ \hbox{PFA}&= \int_R
  f(\zb|{\Theta_v=0})\,d\zb, \label{eq:PFA-R}\end{align} where
$R\subset M$ is the detection region in which observations are
declared to yield the decision ${\Theta_v=1}$, otherwise $\Theta_v$ is
declared to equal~$0$. The optimum Neyman-Pearson test uses the
detection region~$R$ that maximizes PD at a fixed CFAR value
$\PFA_0$. Posing this optimization problem over~$R$ with the method of
Lagrange multipliers applied to the function
\begin{multline} F(R,\lambda) =\PD(R)-\lambda\bigl(\PFA(R)-\PFA_0\bigr),\\ 
=\int_R f(\zb|{\Theta_v=1})\,d\zb -\lambda\Bigl[\int_R
  f(\zb|{\Theta_v=0})\,d\zb-\PFA_0\Bigr]\\ =\int_R\bigl[f(\zb|{\Theta_v=1})-\lambda
  f(\zb|{\Theta_v=0})\bigr]\,d\zb +\lambda\PFA_0\end{multline} yields
two conditions to maximize $F(R,\lambda)$ over $R$
and~$\lambda$: $$\properties{(i)&\lambda&>0,\cr (ii)&z&\in R
  \;\Leftrightarrow\; f(\zb|{\Theta_v=1})-\lambda f(\zb|{\Theta_v=0})
  > 0.\cr}$$ The second property yields the {\em likelihood ratio\/}
(LR) test, \begin{equation} \frac{f(\zb|{\Theta_v=1})}
  {f(\zb|{\Theta_v=0})}\detthreshv{v}\lambda \label{eq:LR}\end{equation}
that maximizes the probability of detection. As will be shown in the
next section, the numerator $f(\zb|{\Theta_v=1})$ of Eq.~(\ref{eq:LR})
is easily computed using standard Bayesian analysis, leading to a
``threat propagation'' algorithm for $f(\Theta_v|\zb)$ and a
connection to the Laplacian~$\LPolishb(G)$ described in
Section~\ref{sec:agt}, and the denominator $f(\zb|{\Theta_v=0})$ is
determined by prior background information or simply the ``principle
of insufficient reason''~\cite{Keynes1921} in which this term is a
constant.

Because the probability of detecting threat is maximized at each
vertex, the probability of detection for the entire subgraph is also
maximized, yielding an optimum Neyman-Pearson test under the
simplification of treating the $2^n$-ary multiple hypothesis testing
problem as a sequence of $n$ binary hypothesis tests. Summarizing, the
probability of network detection given an observation~$\zb$ is
maximized by computing $f(\Theta_v|\zb)$ using a Bayesian ``threat
propagation'' method and applying a simple likelihood ratio test. The
connectivity of the subgraph whose vertices exceed the threshold is
assured by the maximum principle.  Algorithms for computing
$f(\Theta_v|\zb)$ are described next.

\subsection{Space-Time Threat Propagation}
\label{sec:sttp}

Many important network detection applications, especially networks
based on vehicle tracks and computer communication networks, involve
directed graphs in which the edges have departure and arrival times
associated with their initial and terminal vertices. Space-Time threat
propagation is used compute the time-varying threat across a graph
given one or more observations at specific vertices and
times.\cite{Philips2012,Smith2012} In such scenarios, the time-stamped
graph~${G=(V,E)}$ may be viewed as a {\em space-time graph\/} ${G_T
  =(V\times T,E_T)}$ where $T$ is the set of sample times and
$E_T\subset[{V\times T}]^2$ is an edge set determined by the temporal
correlations between vertices at specific times. This edge set is
application-dependent, but must satisfy the two constraints, (1)~if
${\bigl(u(t_k),v(t_l)\bigr)\in E_T}$ then ${(u,v)\in E}$, and
(2)~temporal subgraphs $\bigl((u,v),E_T(u,v)\bigr)$ between any two
vertices $u$ and~$v$ are defined by a temporal model $E_T(u,v)\subset
[{T\coprod T}]^2$. A concrete example for a specific dynamic model of
threat propagation is provided below.

\subsubsection{Temporal Threat Propagation}
\label{sec:ttp}

Given an observed threat at a particular vertex and time, we wish to
compute the inferred threat across all vertices and all times. This
computation is a straightforward application of Bayesian analysis that
results in the optimum Neyman-Pearson network detection test developed
above as well as an efficient algorithm for computing this test. Given
a vertex~$v$, denote the threat at~$v$ and at time~${t\in\Rbbb}$ by
the $\{\,0,1\,\}$-valued stochastic process $\Theta_v(t)$, with value
zero indicating no threat, and value unity indicating a threat. Denote
the\/ {\em probability of threat\/} at~$v$ at~$t$ by
\begin{equation}
  \vartheta_v(t) \buildrel{\rm def}\over= P\bigl(\Theta_v(t)=1\bigr)
  =P\bigl(\Theta_v(t)\bigr). \label{eq:Pthreat}
\end{equation}
The threat state at~$v$ is modeled by a finite-state continuous time
Markov jump process between from state~$1$ to state~$0$ with Poisson
rate~$\lambda_v$. With this simple model the threat stochastic process
$\Theta_v(t)$ satisfies the It\^o stochastic differential equation,
\begin{equation} d\Theta_v =-\Theta_v\,dN_v;\quad
  \Theta_v(0)=\theta_1, \label{eq:sttpIto}
\end{equation}
where $N_v(t)$ is a Poisson process with rate~$\lambda_v$ defined for
positive time, and simple time-reversal provides the model for
negative times.  Given an observed threat ${z=\Theta_v(0)=1}$ at~$v$
at~${t=0}$ so that ${\vartheta_V(0)=1}$, the probability of threat
at~$v$ under the Poisson process model (including time-reversal) is
\begin{equation} \vartheta_v(t)
  =P\bigl(\Theta_v(t)|z=\Theta_v(0)=1\bigr)
  =e^{-\lambda_v|t|}, \label{eq:stprob}
\end{equation}
This stochastic model provides a Bayesian framework for inferring, or
propagating, threat at a vertex over time given threat at a specific
time. The function
\begin{equation}
K_v(t)=e^{-\lambda_v|t|}\label{eq:stkernel}
\end{equation}
of Eq.~(\ref{eq:stprob}) is called the {\em space-time threat kernel}
and when combined with spatial propagation provides a temporal model
$E_T$ for a space-time graph. A Bayesian model for propagating threat
from vertex to vertex will provide a full space-time threat
propagation model and allow for the application of the optimum maximum
likelihood test of Eq.~(\ref{eq:LR}).

\subsubsection{Spatial Threat Propagation}
\label{sec:stp}

Propagation of threat from vertex to vertex is determined by tracks or
connections between vertices. A straightforward Bayesian analysis
yields nonlinear equations that determine the probability of threat at
each vertex, and along with the assumptions of asymptotic independence
and small probabilities these equations may be linearized and thereby
easily analyzed and solved in regimes relevant to our applications.

The threat at vertex~$v$ at which a single track~$\tau$ from
vertex~$u$ arrives and\slash or departs at times $t_{\tau}^v$
and~$t_{\tau}^{u}$ is determined by Eq.~(\ref{eq:stprob}) and the
(independent) event $v\gets u$ that threat traveled along this track:
$P\bigl(\Theta_v(t)\bigr) =\vartheta_v(t)
=\vartheta_{u}(t_{\tau}^{u})\* K_v(t-t_{\tau}^v)\* P(v\gets u)$. There
is a linear transformation
\begin{multline}\vartheta_v(t) =P(v\gets
    u)K(t-t_{\tau}^v)\vartheta_{u}(t_{\tau}^{u})\\ =\int_{-\infty}^\infty
    P(v\gets u) K(t-t_{\tau}^v) \delta(\sigma-t_{\tau}^{u})
    \vartheta_{u}(\sigma)\,d\sigma
\end{multline} from the threat
probability at~$u$ to~$v$. Discretizing time, the temporal matrix
$\Kb_{\tau}^{uv}$ for the discretized operator has the sparse form
\begin{equation} \Kb_{\tau}^{uv}
    =\Bigl(\>\zerob\;\ldots\;\zerob\>
    K(t_k-t_\tau^v)\>\zerob\;\ldots\;\zerob\>\Bigr),\label{eq:Kuv}
\end{equation} where $\zerob$ represents an all-zero column, $t_k$
represents a vector of discretized time, and the discretized function
$K(t_k-t_{\tau}^v)$ appears in the column corresponding to the
discretized time at~$t_{\tau}^{u}$.  Threat propagating from
vertex~$v$ to~$u$ along the same track $\tau$ is given by the
comparable expression $\vartheta_{u}(t) =\vartheta_v(t_{\tau}^v)
K(t-t_{\tau}^{u})$, whose discretized linear operator
$\Kb_{\tau}^{vu}$ takes the form \begin{equation}\Kb_{\tau}^{vu}
  =\Bigl(\>\zerob\;\ldots\;\zerob\>
  K(t_k-t_{\tau}^{u})\>\zerob\;\ldots\;\zerob\>\Bigr)\end{equation}
[cf.\ Eq.~(\ref{eq:Kuv})] where the nonzero column corresponds
to~$t_{\tau}^v$. The sparsity of~$\Kb_{\tau}^{uv}$
and~$\Kb_{\tau}^{vu}$ will be essential for practical space-time
threat propagation algorithms.

It will now be shown how threats arriving on other tracks from other
vertices may be sequentially linearized. If the threat on the
track~$\tau$ from vertex~$u$ must be combined with an existing
threat~$\vartheta_v(t)$ at~$v$, then the combined threat
${\vartheta_v(t^\pm)=P\bigl(\Theta_v(t^\pm)\bigr)}$ at~$v$ at
time~$t^\pm$ immediately after\slash before the track from $u$
arrives\slash departs at time~$t$ is determined by the addition law of
probability,
\begin{multline} P\bigl(\Theta_v(t)\cup
  \Theta_u(t)({v\gets u})\bigr)\\=P\bigl(\Theta_v(t)\bigr)
  +P\bigl(\Theta_u(t)({v\gets
    u})\bigr)\\ {}-P\bigl(\Theta_v(t){\cdot}\Theta_u(t)({v\gets
    u})\bigr). \label{eq:addlaw}
\end{multline}
Under the two assumptions that (1)~the threat events $\Theta_u$ and $\Theta_v$ at~$u$ and~$v$ are independent, asymptotically valid for large time differences relative to the Poisson time $\lambda_{v^*}^{-1}$ for an observation at vertex~$v^*$,\cite{Smith2012} and (2)~the threat probabilities $P\bigl(\Theta_u(t)\bigr)$ and~$P\bigl(\Theta_v(t)\bigr)$ are numerically small, Eq.~(\ref{eq:addlaw}) yields the linear approximation
\begin{multline}
  \vartheta_v(t^\pm) \approx P\bigl(\Theta_v(t)\bigr)
  +P\bigl(\Theta_u(t)({v\gets u})\bigr)\\ {}=\vartheta_v(t)
  +\vartheta_u(t)P({v\gets u}). \label{eq:addapprox}
\end{multline}
Extending this analysis to multiple tracks and assuming that
$P({v\gets u})^{-1}\propto w(v)$ for some weight function $w\colon
V\to\Rbbb$ of the vertices, e.g., the degree of each vertex, yields the {\em threat propagation equation\/}
\begin{equation}
\varthetab
  =\Db^{-1}\Ab\varthetab, \label{eq:tpeqn}
\end{equation}
where $\varthetab$ is the (discretized) space-time vector of threat
probabilities, the weighted space-time adjacency matrix
\begin{equation} \Ab_{uv} =\begin{pmatrix}\zerob
  &\sum_l\Kb_{\tau_l}^{vu}\\ \sum_l\Kb_{\tau_l}^{uv}
  &\zerob\end{pmatrix}
\end{equation}
is defined by Eq.~(\ref{eq:Kuv}), and $\Db^{-1}
=\diag\bigl(w(v_1)\Ib,\ldots,w(v_n)\Ib\bigr)$. Eq.~(\ref{eq:tpeqn}),
written as ${\LPolishb\varthetab=\zerob}$, connects the asymmetric
Laplacian matrix of Eq.~(\ref{eq:genlaplacian}) with threat
propagation, the solution of which itself may be viewed as a boundary
value problem with the harmonic operator~$\LPolishb$.

\begin{figure*}[t]
\medskip
\normalsize
\centerline{\includegraphics[width=0.75\linewidth]{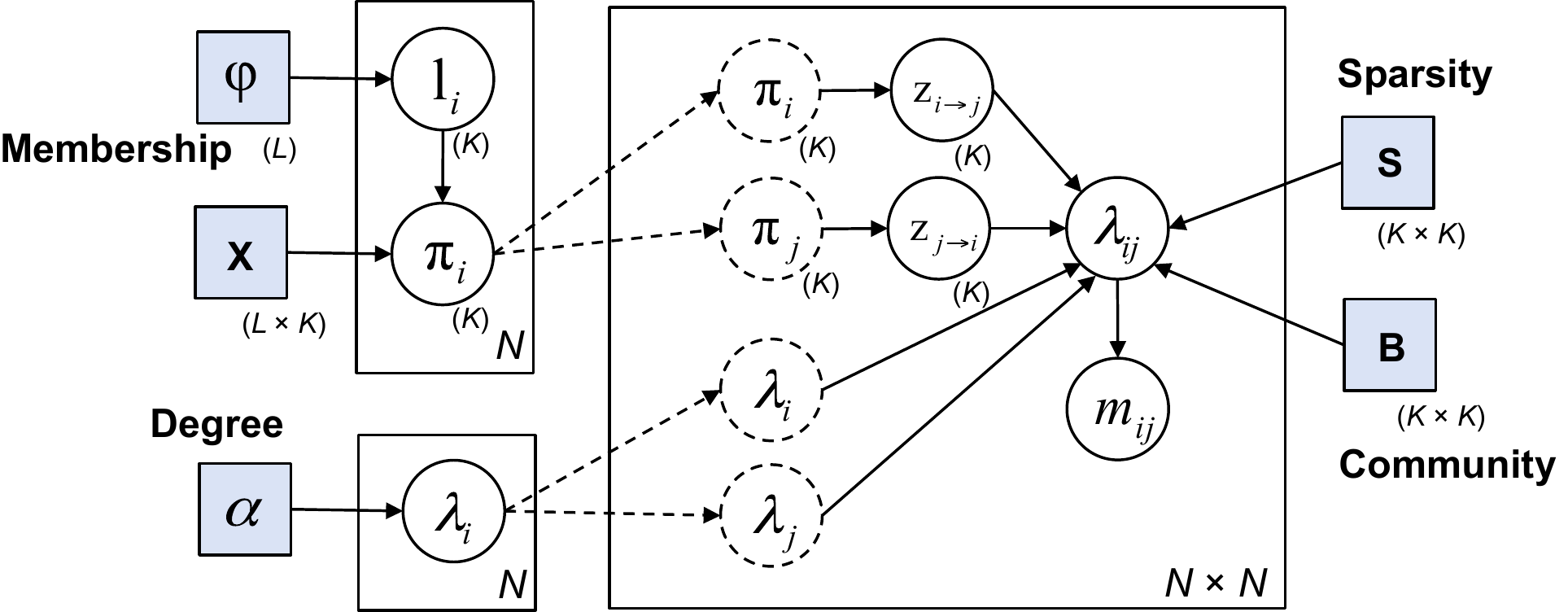}}
\caption{Bayesian generative model for the network simulation with $N$
  nodes, $K$ communities, and $L$ ''lifestyles'' (distributions of
  community participation). Shaded squares are model parameters for
  tuning and circles are variables drawn during
  simulation.\label{fig:blockmodel}}
\end{figure*}

Given a cue at vertices $v_{b_1}$,~\ldots, $v_{b_C}$, the {\em
  harmonic space-time threat propagation equation\/}
is \begin{equation}\textstyle \Bigl(\LPolishb_{\rm ii}\;\LPolishb_{\rm
    ib}\Bigr)\Bigl({\varthetab_{\rm i}\atop \varthetab_{\rm b}}\Bigr)
  =\zerob\label{eq:hsttpe}\end{equation} where the space-time
Laplacian~$\LPolishb= \Bigl({\LPolishb_{\rm ii}\atop \LPolishb_{\rm
    bi}}\, {\LPolishb_{\rm ib}\atop \LPolishb_{\rm bb}}\Bigr)$ and the
space-time threat vector $\varthetab=\Bigl({\varthetab_{\rm i}\atop
  \varthetab_{\rm b}}\Bigr)$ have been permuted so that cued vertices
are in the `\/${\rm b}$' blocks (the ``boundary''), non-cued vertices
are in `\/${\rm i}$' blocks (the ``interior''), and the cued
space-time vector $\varthetab_{\rm b}$ is given. The {\em harmonic
  threat} is the solution to
Eq.~(\ref{eq:hsttpe}), \begin{equation}\varthetab_{\rm i} =
  -\LPolishb_{\rm ii}^{-1} (\LPolishb_{\rm ib}\varthetab_{\rm
    b}). \label{eq:harmonicthreat}\end{equation} The space-time
Laplacian of Eq.~(\ref{eq:genlaplacian}) is a directed Laplacian
matrix, and that Eq.~(\ref{eq:hsttpe}) is directly analogous to
Laplace's equation ${\Delta\varphi=0}$ given a fixed boundary
condition. As discussed in the next subsection, the connection between
space-time threat propagation and harmonic graph analysis also
provides a link to spectral-based methods for network detection.  The
nonnegativity of the harmonic threat of Eq.~(\ref{eq:harmonicthreat})
is guaranteed because the space-time adjacency matrix~$\Ab$ and cued
threat vector $\varthetab_{\rm b}$ are both nonnegative.  This highly
sparse linear system may be solved by the biconjugate gradient method,
which provides a practical computational approach that scales well to
graphs with thousands of vertices and thousands of time samples,
resulting in space-time graphs of order ten~million or more. In
practice, significantly smaller subgraphs are encountered in
applications such as threat network discovery~\cite{Smith2011}, for
which linear solvers with sparse systems are extremely fast.

Finally, a simple application of Bayes' theorem to the harmonic threat
${\vartheta_v=f(\Theta_v|\zb)}$ provides the optimum Neyman-Pearson
detector [Eq.~(\ref{eq:LR})] developed in Section~\ref{sec:npnt}
because \begin{multline}
  \frac{f(\zb|{\Theta_v=1})}{f(\zb|{\Theta_v=0})}
  =\frac{f({\Theta_v=1}|\zb)}{f({\Theta_v=0}|\zb)}
  \cdot\frac{f({\Theta_v=0})}{f({\Theta_v=1})}\\ =\frac{\vartheta_v}{f({\Theta_v=0}|\zb)}
  \cdot\frac{f({\Theta_v=1})}{f({\Theta_v=0})}
  \detthreshv{v}\lambda, \label{eq:LRsttp} \end{multline} results in a
threshold of the harmonic space-time threat propagation
vector \begin{equation}\varthetab
  \detthresh\mbox{threshold}, \label{eq:sttpthresh}\end{equation}
possibly weighted by a nonuniform null distribution
$f({\Theta_v=0}|\zb)$, with the normalizing constant
$f({\Theta_v=1})/f({\Theta_v=0})$ being absorbed into the detection
threshold. This establishes, under the assumptions and approximations
enumerated above, the detection optimality of harmonic space-time
threat propagation.

\subsection{Insights from Spectral Graph Theory}
\label{sec:connect}

Each network detection algorithm above can be compared to each other
by different approaches taken to address the problem posed by the
(physical) fact that the smallest eigenvalue of the graph Laplacian is
zero: ${\Qb\oneb=0{\cdot}\oneb}$. Fiedler's spectral detection, which
minimizes the network cut size, thresholds the eigenvector
corresponding to the second smallest eigenvalue of the Laplacian---the
Fiedler value. In contrast, community detection, which maximizes the
subgraph connectivity relative to the background, recasts the
objective of spectral detection resulting in a threshold of the
principal or other eigenvectors of Newman's modularity matrix
$\Mb=\Ab-V^{-1}\db\db^\T$. Alternatively, threat propagation, which
maximizes the Bayesian probability of detection by computing the
harmonic solution to Laplace's equation,
${\LPolishb\varthetab=\zerob}$, but treats this as a boundary value
problem with observations representing the boundary values and unknown
values representing the interior.

\subsection{Computational Complexity}
\label{sec:complexity}

Depending upon sparsity, the computational complexity of spectral
methods ranges from~$O(n\log n)$--$O(n^2)$ for principal eigenvector
methods~\cite{Newman2006} to~$O(n^2\log n)$--$O(n^3)$ for methods that
rely on full eigensolvers~\cite{Miller2010a,Miller2010b,Miller2011},
with the lower cost exhibited with graphs whose average degree is
over~$\log n$, below which a random Erd\humlaut os-R\'enyi graph is
almost surely disconnected.\cite{Erdos1960} The cost of harmonic
methods is about $O(n\log n)$--$O(n^2)$ for sparse matrix inversion
and also depends upon the graph's sparsity. In practice, Arnoldi
iteration can be used for sparse eigenvalue computation and the
biconjugate gradient method can be used for sparse matrix inversion.

\section{Network Detection Performance}
\label{sec:netdetperf}

There are two ways to demonstrate network detection performance:
empirical and theoretical, both of which depend on detailed knowledge
of network behavior and dynamics.  But full knowledge of real-world
covert network behavior including relationships to the background
network is, by design, extraordinarily rare or nonexistent, though
partial information about many covert networks has been integrated
over time~\cite{Xu2008}. Predicting performance of network detection
methods requires details of the interconnectivity of both the
foreground and background networks. Empirical detection performance is
demonstrated using either a real-world or simulated dataset for which
the truth is at~least partially known, and theoretical performance
predictions are derived based upon statistical assumptions about the
foreground and background networks. To~date, closed-form analytic
performance predictions have been accomplished for very simple network
models, i.e.\ cliques~\cite{Fortunato2007,Kumpula2007,Nadakuditi2012}
or dense subgraphs~\cite{Arias-Castro2013} embedded within
Erd\humlaut{o}s-R\'enyi backgrounds, and there are no theoretical
results at~all for space-time graphs or realistic models appropriate
for covert networks. Therefore, realistic models are essential for
performance analysis of network detection algorithms. There are two
basic approaches to modeling networks: stochastic models, which
attempt to capture the aggregate statistical properties of networks,
and agent-based models, which attempt to describe specific
behaviors. In general, stochastic models have greater tractability
because they do not rely on the detailed description of actions or
objectives of a specific network.

The empirical detection performance of the covert network detection
algorithms described above will be computed using a Monte-Carlo
analysis based upon a new stochastic blockmodel. Empirical performance
predictions may be also based on a single dataset, oftentimes a
practical necessity for real-world measurements. Detection performance
for specific, real-world single datasets is illustrated in an
accompanying paper.\cite{Yee2012}

\subsection{Covert Network Stochastic Blockmodel}
\label{sec:covnetsb}

To adhere with observed phenomenology of real-world networks,
realistic network models should exhibit properties including
connectedness, a power-law degree distribution (the ``small world''
property), membership-based community structure, sparsity, and
temporal coordination.  No one simple network model captures all these
traits, e.g.\ Erd\humlaut{o}s-R\'enyi graphs can be almost surely
connected, though do not exhibit a power-law density, power-law models
such as R-MAT\cite{Chakrabarti2004} do not exhibit a membership-based
network structure, and mixed-membership stochastic
blockmodels~\cite{Airoldi2008} do not include temporal
coordination. To achieve a realistic network model possessing this
range of properties, we propose a new statistical-based model with
parameterized control over the generation of interactions between
network nodes.  The proposed model is depicted in
Fig.~\ref{fig:blockmodel} using plate notation.

\begin{figure}[t]
\medskip
\normalsize
\centerline{\includegraphics[width=0.75\linewidth]{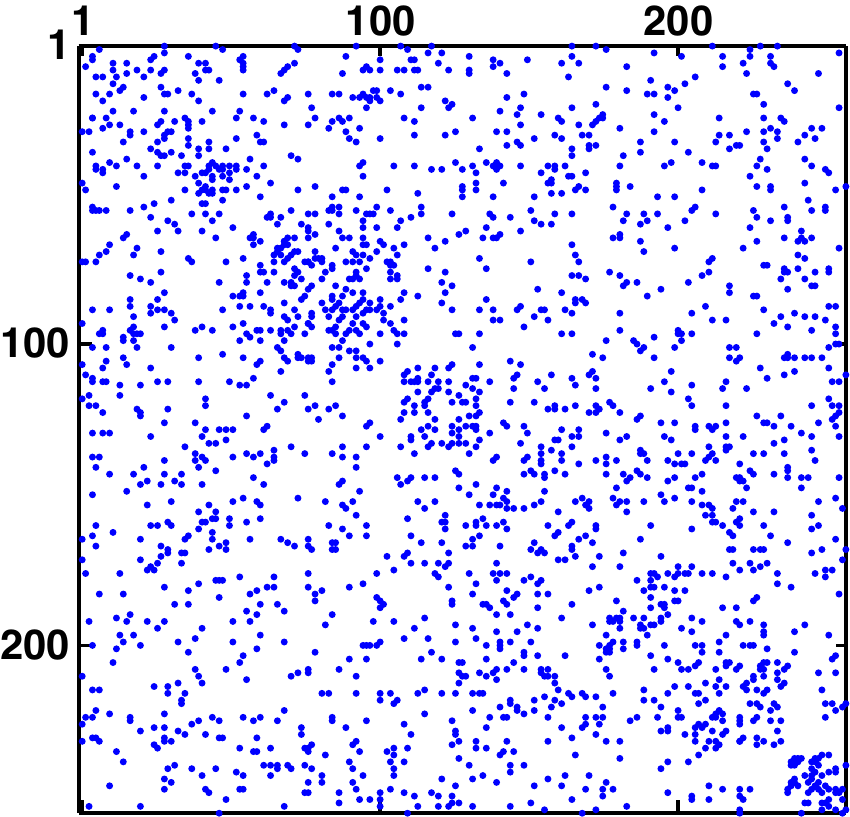}}
\caption{Adjacency matrix of a stochastic blockmodel with a
  foreground subgraph whose intra-activity is 50\%\ more than of all
  other subgraphs.\label{fig:adjacency_matrix}}
\end{figure}

\subsubsection{Spatial Stochastic Blockmodel}
\label{sec:spatial}

The proposed model may be viewed as an aggregation of the several
simpler models of which it is comprised: Erd\humlaut{o}s-R\'enyi
(dominant at low degrees),~\cite{Erdos1960} Chung-Lu (dominant at high
degrees),~\cite{Aiello2001} and a mixed-membership blockmodel that
models community interactions.~\cite{Airoldi2008} The overall network
model is approximated by each of the simpler models in the regime
where the simple model dominates.  The Erd\humlaut{o}s-R\'enyi model
defines the overall sparsity and connectivity.  The Chung-Lu model
creates a power-law degree distribution empirically consistent with a
broad range of real-world networks. The stochastic blockmodel creates
distinct communities each with their own parameterized interaction
models.

The space-time graph of the proposed mixed-membership stochastic
blockmodel is determined by a connectivity model and temporal
model. Let $N$ be the total number of nodes, and $K$ be the number of
communities. Each node divides its time among at least one of the
several $K$ communities, and the number of ways in which a node
distributes its time among the different communities is discretized
into $L$ distinct ``lifestyles.'' Each node is assigned to a specific
lifestyle. For example, nodes~$1$ and $3$ may spend all their time in
community~$1$, thereby sharing the same lifestyle, whereas node~$2$
may spend half its time in community~$1$ and half in community~$2$,
and therefore occupies another lifestyle, and so forth.  The
rate~$\lambda_{ij}$ of interactions between nodes~$i=1$ and~$j$ is
given by the product
\begin{equation}\lambda_{ij} =I_{ij}^\Sb\cdot
  \frac{\lambda_i\lambda_j} {\sum_k\lambda_k}\cdot\zb_{i\rightarrow
    j}^\T \Bb \zb_{i\rightarrow j} ,\label{eq:bmrates}
\end{equation}
where the first term $I_{ij}^\Sb$ represents the (modified)
Erd\humlaut{o}s-R\'enyi model, the second term
$\lambda_i\lambda_j/\bigl(\sum_k\lambda_k\bigr)$ represents the
Chung-Lu model, and the third term $\zb_{i\rightarrow j}^\T \Bb
\zb_{i\rightarrow j}$ represents the stochastic blockmodel.

\begin{figure}[t]
\medskip
\normalsize
\centerline{\includegraphics[width=0.7083\linewidth]{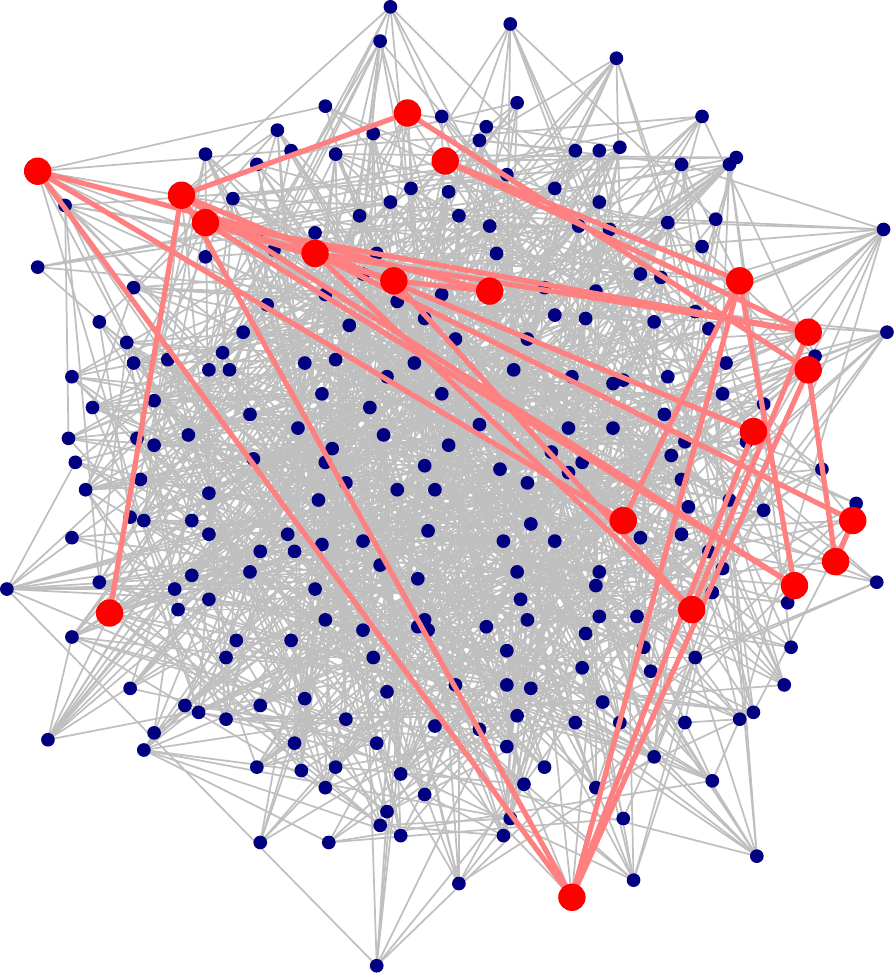}}
\caption{Graph of the adjacency matrix shown in
  Figure~\ref{fig:adjacency_matrix}. The foreground graph and intra
  subgraph edges are shown in
  {\color{red}red}.\label{fig:graphviz256}}
\end{figure}

At each node-to-node interaction, a random draw from a multinomial
distribution determines the community to which each node belongs.  The
indicator function $I_{ij}^\Sb$ is a sparse $K\by K$ $(0,1)$-matrix
whose entries are binomial random variables with probability
$(\Sb)_{ab}$ for node~$i$ in community~$a$ and node~$j$ in
community~$b$. An Erd\humlaut{o}s-R\'enyi sparsity model has
$(\Sb)_{ab}\equiv p$ for~all communities $a$ and~$b$, whereas this
modified sparsity model allows the possible of differing interaction
rates within and across communities.  The Chung-Lu term
$\lambda_i\lambda_j/\bigl(\sum_k\lambda_k\bigr)$ is determined by the
per-node expected degrees $\lambda_i$, $i=1$, \dots,~$N$, which are
themselves drawn from a power-law distribution of parameter
${\alphab\in\Rbbb^N}$. The blockmodel term $\zb_{i\rightarrow j}^\T
\Bb \zb_{i\rightarrow j}$ is determined by~$\Bb$, a $K\by K$ matrix of
the rate of interaction between communities, and ${\zb_{i\rightarrow
    j}\in\Rbbb^K}$ is an indicator $(01)$-vector is the community to
which node~$i$ belongs when interacting with node~$j$. This community
is the same over the entire simulation and is drawn from a multinomial
over ${\pib_i\in\Rbbb^K}$, node~$i$'s distribution over
communities. Finally, the distribution of~$\pib$ is drawn from a
Dirichlet r.v.\ with concentration parameter $\lb_i^\T\Xb$. Node~$i$'s
lifestyle, $\lb_i$ is a multinomial draw with the lifestyle
probability ${\phib\in\Rbbb^L}$. The adjacency matrix and graph of
this model are illustrated in Figs.\ \ref{fig:adjacency_matrix} and
~\ref{fig:graphviz256} using an example with a mixed community with a
higher level of activity for the foreground network.

\subsubsection{Temporal Stochastic Blockmodel}
\label{sec:temporal}

The meeting times for each interaction are chosen independently of the
spatial model. Real-world interactions are often coordinated, with
many individuals arriving or leaving from a location at a set of
pre-defined times. This behavior is parameterized by an average number
of meeting times ${\Psib\in\Rbbb^K}$ for~each community. The simulated
number of meeting times is a Poisson r.v.\ (offset by~$1$) with
Poisson parameter~${\Psib-1}$. E.g.\ an expected number of meeting
times ${(\Psib)_k=1}$ (for community~$k$) yields a constant Poisson
r.v.\ of~$1$ meeting time (in Matlab, {\tt poissrnd(0) = 0}), thereby
yielding a community whose activities are tightly coordinated because
there is only a single time for the members to meet. An expected
number of meeting times $(\Psib)_k=20$ yields a community whose
activities are loosely coordinated because meetings may occur at any
one of a number of times.
The meetings times themselves are chosen uniformly over time, and each
node arrives at the meeting time perturbed by a zero-mean Gaussian
r.v.\ with a parameterized variance.

\begin{figure}[t]
\medskip
\normalsize
\centerline{\includegraphics[width=0.8333\linewidth]{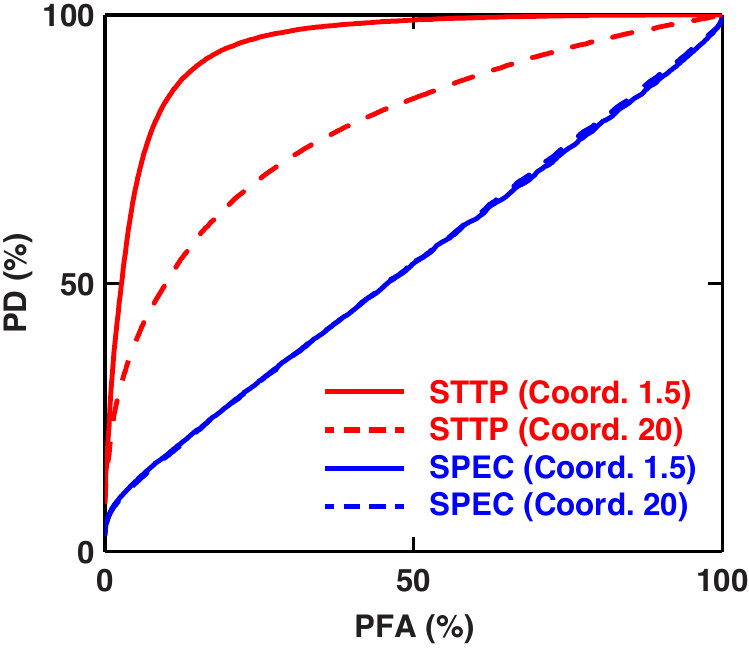}}
\caption{Receiver operating characteristics versus forground
  coordination ($\Psib_{\rm foreground}=1.5$, high coordination, and
  $20$, low coordination) for space-time threat propagation (STTP) and
  spectral-based community detection (SPEC). The community activity
  level $\Sb_{\rm k}= 1{\cdot}\log N_{\rm k}/N_{\rm k}$ for all
  communities. [$1000$ Monte~Carlo trials.]\label{fig:coordroc}}
\end{figure}

\subsection{Network Detection Results}
\label{sec:netdetres}

The detection performance of the network detection algorithms
described above is presented in this section using empirical
Monte~Carlo results applied to the mixed-membership stochastic
blockmodel. A space-time graph is chosen independently for each
Monte~Carlo trial. A set of baseline parameters is chosen to achieve
realistic foreground and background networks of specific sizes, and
excursions are performed on the parameters controlling foreground
coordination and foreground activity. The performance metric is the
standard receiver operating characteristic (ROC), which in the case of
network detection is the probability of detection (measured as the
percentage of true foreground nodes detected) versus the number or
percentage of false alarms (the number of background nodes detected)
as the detection threshold is varied. Perfect ROC performance is a
100\%\ detection rate with a 0\% false alarm rate, and the worst
possible performance is a detection rate equal to chance, i.e.\ equal
to the false alarm rate.

\begin{figure}[t]
\medskip
\normalsize
\centerline{\includegraphics[width=0.8333\linewidth]{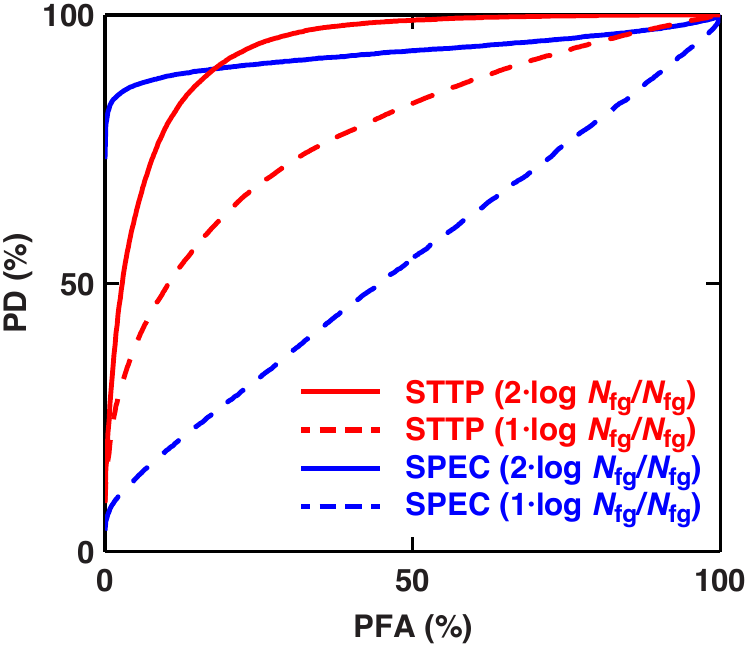}}
\caption{Receiver operating characteristics versus foreground activity
  ($\Sb_{\rm fg}=2{\cdot}\log N_{\rm fg}/N_{\rm fg}$, high activity,
  and $1{\cdot}\log N_{\rm fg}/N_{\rm fg}$, baseline activity) for
  space-time threat propagation (STTP) and spectral-based community
  detection (SPEC). The foreground coordination level is specified
  by~$\Psib_{\rm fg}=20\,$average number of meeting times. [$1000$
    Monte~Carlo trials.]\label{fig:activityroc}}
\end{figure}

\subsubsection{Baseline Model}

A baseline model is used comprised of eleven lifestyles spanning ten
communities. Two of the lifestyles are designated as foreground
lifestyles and all others are ``background.''  As detail above, each
lifestyle has a propensity toward a different mix of community
activity. The background lifestyles have a power-law distribution of
membership over the background communities, which may be imagined to
represent business, homes, industry, religion, sports, or other social
interactions. Two distinct foreground lifestyles are used to model the
compartmentalization of real-world covert networks.  One foreground
lifestyle associates uniformly across background communities, whereas
the other foreground lifestyle has a strong association with a only
small subset of background communities.  These foreground lifestyles
may be imagined to represent specialized functions or activities
within the covert network. As in real life, the foreground lifestyles
comprise only a tiny fraction of the entire population.

Interactions in which two nodes belong to the same community occur at
a higher rate than interactions of nodes belonging to different
communities. This is modeled by specifying that the block matrix~$\Bb$
be diagonally dominant, perhaps strongly. Furthermore, real-world
communities are not disconnected, thus for a community size of~$N_k$,
the diagonals of the Erd\humlaut os-R\'enyi sparsity parameter matrix
$\Sb_k$ must be at least $\log N_k/N_k$ to ensure that each community
is almost surelyconnected~\cite{Erdos1960}. Finally, covert networks
necessarily have sparse---not clique-like---structure, thus the
Erd\humlaut os-R\'enyi sparsity parameter $\Sb$ for the covert network
must also be low.

\subsubsection{Detection versus Foreground Coordination and Activity}

Two nominal values are chosen for foreground coordination and
foreground activity, then both space-time threat propagation and
spectral-based community detection algorithms are applied using a
randomized cue over $1000$ Monte~Carlo trials.
Fig.~\ref{fig:coordroc} shows the detection performance of both
algorithms as the foreground coordination changes from a high of
${\Psib_{\rm fg}=1.5}\,$average number of meeting times to a low of
${\Psib_{\rm fg}=20}$. As predicted, the detection performance of
space-time threat propagation improves as the temporal coordination of
the foreground network increases. The optimality of this Bayesian
network detector is predicated on temporal coordination, and decreased
coordination makes the foreground network more difficult to
detect. This example uses a constant baseline level of community
activity (sparsity matrix $\Sb_{\rm fg}=1{\cdot}\log N_{\rm fg}/N_{\rm
  fg}$), thus the optimality assumption of high foreground activity
made by spectral-based community detection algorithm is violated, and
as expected this spectral algorithm does no better than chance for
either coordination level. Fig.~\ref{fig:activityroc} shows the
detection performance of both algorithms as the foreground activity
changes from $\Sb_{\rm fg}=1{\cdot}\log N_{\rm fg}/N_{\rm fg}$
(baseline activity) to $\Sb_{\rm fg}=2{\cdot}\log N_{\rm fg}/N_{\rm
  fg}$ (high activity). The foreground coordination level is low, at
$\Psib_{\rm fg}=20$, providing an example for which none of the basic
algorithmic assumptions hold for either space-time threat propagation
or spectral-based community detection. The low foreground activity
results, $\Sb_{\rm fg}=1{\cdot}\log N_{\rm fg}/N_{\rm fg}$, are
replicated in this figure from Fig.~\ref{fig:coordroc}, in which STTP
yields moderate detection performance and spectral-based community
detection is no better than chance.  At high foreground activity the
foreground network is detectable at by both spectral-based community
detection and space-time threat propagation.

\section{Conclusions}
\label{sec:conc}

The problem of covert network detection is analyzed from the
perspectives of graph partitioning and algebraic graph theory. Network
detection is addressed as a special case of graph partitioning in
which membership in a small subgraph of interest must be determined,
and a common framework is developed to analyze and compare different
network detection methods. A new Bayesian network detection framework
called space-time threat propagation is introduced that partitions the
graph based on prior information and direct observations. Space-time
threat propagation is shown to be optimum in the Neyman-Pearson sense
subject to the assumption that threat networks are connected by edges
temporally correlated to a cue or observation. Bayesian space-time
threat propagation is interpreted as the solution to a harmonic
boundary value problem on the graph, in which a linear approximation
to Bayes' rule determines determines the unknown probability of threat
on the uncued nodes (the ``interior'') based on threat observations at
cue nodes (the ``boundary''). This new method is compared to
well-known spectral methods by examining competing notions of network
detection optimality. Finally, a new generative mixed-membership
stochastic blockmodel is introduced for performance prediction network
detection algorithms. The parameterized model combines key real-world
aspects of several random graph models: Erd\humlaut os-R\'enyi for
sparsity and connectivity, Chung-Lu for power-law degree
distributions, and a mixed-membership stochastic blockmodel for
distinctive community-based interaction and dynamics. This model is
used to compute empirical detection performance results for the
detection algorithms described in the paper as both foreground
coordination and activity levels are varied.  Though the results in
the paper are empirical, it is our hope that both the paper's analytic
results and performance modeling will be useful in future closed-form
analysis of real-world covert network detection problems.

\let\bibliographysize=\footnotesize 

\bibliography{strings,refs}

\begin{thebibliography}{{\bibliographysize99}}

\def\thinskip{\hskip .16667em }

{\catcode`\@=11 \catcode`|=\active \catcode`\!=\active
\gdef\references{\catcode`|=\active \catcode`\!=\active
  \def\!{\char`\!}%
  \def|##1|{{\sc ##1}}      
  \def!##1!{\emph{##1}}     
  \def\<##1>{{\bf##1}\futurelet\next\number@ptional}  
  \def\number@ptional{\ifx\next(\def\@temp{\n@mber}\else
    \ifx\next:\def\@temp{\p@ges}\else\def\@temp{}\fi\fi \@temp}%
  \def\n@mber(##1){\thinspace(##1)\futurelet\next\page@ptional}%
  \def\page@ptional{\ifx\next:\def\@temp{\p@ges}\else
    \def\@temp{}\fi \@temp}%
  \def\p@ges:{\thinspace:\penalty-20\thinskip}%
  \frenchspacing}}
\def\authorbar{$\vcenter{\hbox{\vrule width3em height.4pt}}\,$}

\def\ieeev#1-{{#1}\hbox{-}\nobreak}

\begingroup

\bibliographysize
\baselineskip=9pt
\parskip=2pt plus 1pt minus1pt
\itemsep=0pt
\interlinepenalty=1000  
\hbadness=1500          
\references


\bibitem{Aiello2001} |W.~Aiello|, |F.~Chung|, and |L.~Lu|. ``A random
  graph model for power law graphs,'' !Experimental Mathematics\/!
  \<10>(1):53--66 (2001).

\bibitem{Airoldi2008} |E.~M. Airoldi|, |D.~M. Blei|, |S.~E. Fienberg|,
  and |E.~P. Xing|. ``Mixed-membership stochastic blockmodels,''
  !JMLR\/! \<9>:1981--2014 (2008).

\bibitem{Alanyali2004} |M.~Alanyali|, |S.~Venkatesh|, |O.~Savas|, and
  |S.~Aeron|. ``Distributed Bayesian hypothesis testing in sensor
  networks,'' in !Proc. 2005 American Control Conf!. Boston~MA,
  pp.~5369--5374 (2004).

\bibitem{Arias-Castro2013} |E.~Arias-Castro| and
  |N.~Verzelen|. ``Community Detection in Random Networks,'' {\tt
    arXiv:1302.7099 [math.ST]}. Accessed 18~March 2013.
  $\langle$\href{http://arxiv.org/abs/1302.7099}{http://arxiv.org/\discretionary{}{}{}abs/1302.7099}$\rangle$.

\bibitem{Belkin2003} |M.~Belkin| and |P.~Niyogi|. ``Laplacian
  eigenmaps for dimensionality reduction and data representation,''
  !Neural Computation\/! \<15>:1373--1396 (2003).

\bibitem{Brun2004} |A.~Brun|, |H.~Knutsson|, |H.~J. Park|,
  |M.~E. Shenton|, and |C.-F. Westin|. ``Clustering fiber tracts using
  normalized cuts,'' in !Proc. Medical Image Computing and
  Computer-Assisted Intervention (MICCAI 04)\/! (2004). Accessed
  3~September 2012.
  $\langle$\href{http://lmi.bwh.harvard.edu/papers/papers/brunMICCAI04.html}{http://lmi.bwh.harvard.edu/\discretionary{}{}{}papers/papers/\discretionary{}{}{}brunMICCAI04.html}$\rangle$.

\bibitem{Carley2004} |K.~Carley|. ``Estimating vulnerabilities in
  large covert networks,'' in !Proc. 16th Intl. Symp. Command
  and Control Research and Tech. (ICCRTS)!. (San Diego, CA) (2004).

\bibitem{Carter2010} |K.~M. Carter|, |R.~Raich|, and
  |A.~O. Hero~III|. ``On Local Intrinsic Dimension Estimation and Its
  Applications,'' !IEEE Trans. Signal Processing\/!
  \<58>(2):650--663 (2010).

\bibitem{Chakrabarti2004} |D.~Chakrabarti|, |Y.~Zhan|, and
  |C.~Faloutsos|. ``R-MAT: A Recursive Model for Graph Mining,'' in
  !Proc. 2004 SIAM Intl. Conf. Data Mining!. (2004).

\bibitem{Chamberland2003} |J.-F. Chamberland| and
  |V.~V. Veeravalli|. ``Decentralized detection in sensor networks,''
  !IEEE Trans. Signal Processing\/!  \<51>(2):407--416 (2003).

\bibitem{Chung1994} |F.~R.~K. Chung|. !Spectral Graph Theory!,
  Regional Conference Series in Mathematics \<92>.  Providence, RI:
  American Mathematical Society (1994).

\bibitem{Costa2004} |J.~A. Costa| and |A.~O. Hero~III|. ``Geodesic
  entropic graphs for dimension and entropy estimation in manifold
  learning,'' !IEEE Trans. Signal Processing\/! \<52>(8):2210--2221
  (2004).

\bibitem{Diestel2000} |R.~Diestel|. !Graph Theory!. New~York:
  Springer-Verlag, Inc. (2000).

\bibitem{DrWeb2012} |Doctor Web|. ``Doctor Web exposes
  550\thinspace000 strong Mac botnet'', 4~April 2012, accessed
  3~September 2012
  $\langle$\href{http://news.drweb.com/show/?i=2341}{http://news.drweb.com/show/?i=2341}$\rangle$.

\bibitem{Donath1973} |W.~E. Donath| and |A.~J. Hoffman|. ``Lower bounds
  for the partitioning of graphs,'' !IBM J. Res. Development\/!
  \<17>:420--425 (1973).

\bibitem{Erdos1960} |P.~Erd\humlaut os| and |A.~R\'enyi|, ``On the evolution
  of random graphs,'' !Pubs. Mathematical Institute of the Hungarian
  Academy of Sciences\/! \<5>:17--61 (1960).

\bibitem{Ferry2009} |J.~P. Ferry|, |D.~Lo|, |S~.T. Ahearn|, and
  |A.~M. Phillips|. ``Network detection theory,'' in !Mathematical
  Methods in Counterterrorism\/!, eds.\ |N. Memon| et~al.,
  pp.~161--181, Vienna: Springer (2009).

\bibitem{Fiedler1973} |M.~Fiedler|. ``Algebraic connectivity of
  graphs,'' !Czech. Math. J.\/! \<23>(2):298--305 (1973).

\bibitem{Fiedler1975} \authorbar. ``A property of eigenvectors of
  non-negative symmetric matrices and its application to graph
  theory,'' !Czech. Math. J.! \<25>:619--633 (1975).

\bibitem{Fortunato2007} |S.~Fortunato| and
  |M.~Barth\'elemy|. ``Resolution limit in community detection,''
  !PNAS\/! \<104>(1):36--41 (2007).

\bibitem{Fortunato2010} |S.~Fortunato|. ``Community detection in
  graphs,'' !Physics Reports\/! \<486>:75--174 (2010).


\bibitem{Godsil2001} |C.~Godsil| and |G.~Royle|. !Algebraic
  Graph Theory!. New~York: Springer-Verlag, Inc. (2001).

\bibitem{Google2002} |Google|. ``The technology behind Google's great
  results,'' Accessed 3~September 2012
  $\langle$\href{http://www.google.com/onceuponatime/technology/pigeonrank.html}{http://www.google.com/\discretionary{}{}{}onceuponatime/\discretionary{}{}{}technology/\discretionary{}{}{}pigeonrank.html}$\rangle$.

\bibitem{Guruharsha2011} |K.~G. Guruharsha| et~al. ``A protein complex
  network of {\em Drosophila melanogaster},'' !Cell\/!
  \<147>(3):690--703 (2011). Accessed 3~September 2012
  $\langle$\href{http://www.sciencedirect.com/science/article/pii/S0092867411010804}{http://www.sciencedirect.com/\discretionary{}{}{}science/article/pii/\discretionary{}{}{}S0092867411010804}$\rangle$.

\bibitem{Jackson2008} |M.~O. Jackson|. !Social and Economic Networks!,
  Princeton~U. Press (2008).

\bibitem{Jonas2006} |J.~Jonas| and |J.~Harper|. ``Effective
  counterterrorism and the limited role of predictive data mining,''
  !Policy Analysis\/! \<584>. Cato Institute (2006).

\bibitem{Kar2008} |S.~Kar|, |S.~Aldosari|, and
  |J.~F. Moura|. ``Topology for distributed inference on graphs,''
  !IEEE Trans. Signal Processing\/! \<56>(6):2609--2613 (2008).

\bibitem{Keynes1921} |J.~M. Keynes|. !A Treatise on
  Probability!. London: Macmillan and Co.\ (1921).

\bibitem{Koller2009} |D.~Koller| and |N.~Friedman|. !Probabilistic
  Graphical Models!. Cambridge, MA: MIT Press (2009).

\bibitem{Krebs2002} |V.~E. Krebs|. ``Uncloaking terrorist networks,''
  !First Monday\/! \<7>(4) (2002).

\bibitem{Kumpula2007} |J.~M. Kumpula|, |J.~Saram\"aki|, |K.~Kaski|,
  and |J.~Kert\'esz|. ``Limited resolution in complex network
  community detection with Potts model approach,'' !Eur. Phys. J. B!
  \<56>:41--45 (2007).

\bibitem{Leskovec2010} |J.~Leskovec|, |K.~J. Lang|, and
  |M.~Mahoney|. ``Empirical comparison of algorithms for network
  community detection,'' in !Proc. 19th Intl. Conf. World Wide Web
  (WWW'10)!. Raleigh, NC, pp.~631--640 (2010).


\bibitem{Miller2011} |B.~A. Miller|, |M.~S. Beard|, and
  |N.~T. Bliss|. ``Eigenspace Analysis for Threat Detection in Social
  Networks,'' in !Proc.\ 14th Intl.\ Conf.\ Informat.\ Fusion
  (FUSION)!. Chicago, IL (2011).

\bibitem{Miller2010a} |B.~A. Miller|, |N.~T. Bliss|, and
  |P.~J. Wolfe|. ``Toward signal processing theory for graphs and
  other non-Euclidean data,'' in !Proc. IEEE Intl. Conf. Acoustics,
  Speech and Signal Processing!, pp.~5414--5417 (2010).

\bibitem{Miller2010b} \authorbar.
  ``\href{http://books.nips.cc/papers/files/nips23/NIPS2010_0954.pdf}
  {Subgraph detection using eigenvector $L_1$ norms},'' in
  !Proc.\ 2010 Neural Information Processing Systems
  (NIPS)!. Vancouver, Canada (2010).

\bibitem{Mohar1991} |B.~Mohar|. ``The Laplacian Spectrum of Graphs,''
  in !Graph Theory, Combinatorics, and Applications!, \<2>,
  eds.\ |Y.~Alavi|, |G.~Chartrand|, |O.~R. Oellermann|, and
  |A.~J. Schwenk|. New~York: Wiley, pp.~871--898 (1991).

\bibitem{Nadakuditi2012} |R.~R. Nadakuditi| and
  |M.~E.~J. Newman|. ``Graph spectra and the detectability of
  community structure in networks,'' !Phys. Rev. Lett.! \<108>, 188701
  (2012).


\bibitem{Newman2006} |M.~E.~J. Newman|. ``Finding community structure in
  networks using the eigenvectors of matrices,'' !Phys. Rev. E!,
  \<74>(3) (2006).

\bibitem{Newman2010} \authorbar. !Networks: An Introduction!.p Oxford
  U. Press (2010).

\bibitem{Philips2012} |S.~Philips|, |E.~K. Kao|, |M.~Yee|, and
  |C.~C. Anderson|. ``Detecting activity-based communities using dynamic
  membership propagation,'' in !Proc. IEEE Intl. Conf. Acoustics,
  Speech and Signal Processing (ICASSP)!. Kyoto, Japan (2012).

\bibitem{Perry2008} |J.~W. Perry| et al. !Protecting Individual
  Privacy in the Struggle Against Terrorists: A Framework for Program
  Assessment!. The National Academies Press (2008).

\bibitem{Pothen1990} |A.~Pothen|, |H.~Simon|, and
  |K.-P. Liou|. ``Partitioning sparse matrices with eigenvectors of
  graphs,'' !SIAM J. Matrix Anal.\ Appl.! \<11>:430--45 (1990).

\bibitem{Sageman2004} |M.~Sageman|. !Understanding Terror Networks!.
  Philadelphia, PA: U.~Pennsylvania Press (2004).

\bibitem{Shamir2008} |A.~Shamir|. ``A survey on mesh segmentation
  techniques,'' !Computer Graphics Forum\/! \<27>(6):1539--1556
  (2008). Accessed 3~September 2012
  $\langle$\href{http://www.faculty.idc.ac.il/arik/site/mesh-segment.asp}{http://www.faculty.idc.ac.il/\discretionary{}{}{}arik/site/\discretionary{}{}{}mesh-segment.asp}$\rangle$.

\bibitem{Shi2000} |J.~Shi| and |J.~Malik|. ``Normalized cuts and image
  segmentation,'' !IEEE Trans. Pattern Anal. Mach. Intell.!
  \<22>(8):888-905 (2000).

\bibitem{Smith2011} |S.~T. Smith|, |A. Silberfarb|, |S. Philips|,
  |E.~K. Kao|, and |C.~C. Anderson|. ``Network Discovery Using Wide-Area
  Surveillance Data,'' in !Proc.\ 14th Intl.\ Conf.\ Informat.\ Fusion
  (FUSION)!. Chicago, IL (2011).

\bibitem{Smith2012} |S.~T. Smith|, |S.~Philips|, and
  |E.~K. Kao|. ``Harmonic space-time threat propagation for graph
  detection,'' in !Proc. IEEE Intl. Conf. Acoustics, Speech and Signal
  Processing (ICASSP)!. Kyoto, Japan (2012).

\bibitem{TeleGeography2010} |TeleGeography|. ``Global Traffic Map
  2010,'' PriMetrica, Inc. Accessed 3~September 2012
  $\langle$\href{http://www.telegeography.com/telecom-maps/global-traffic-map/index.html}{http://www.telegeography.com/telecom-maps/global-traffic-map/index.html}$\rangle$.

\bibitem{Trinquier2006} |R.~Trinquier|. !Modern Warfare: A French View
  of Counterinsurgency!. Westport, CT: Praeger Security International
  (2006).

\bibitem{USArmy2006} |United States Army|.  !Counterinsurgency: Field
  Manual 3-24!, Appendix~B. Washington: Government Printing Office
  (2006).

\bibitem{VanTrees1968} |H.~L. Van~Trees|. !Detection, Estimation, and
  Modulation Theory!, Part~1. New~York: John Wiley and Sons,
  Inc. (1968).

\bibitem{vonLuxburg2005} |U.~von~Luxburg|, |O.~Bousquet|, and
  |M.~Belkin|. ``Limits of spectral clustering,'' in !Advances in
  Neural Information Processing Systems\/! \<17>, eds. |L.~K. Saul|,
  |Y.~Weiss|, and |L. Bottou|. Cambridge, MA: MIT Press (2005).

\bibitem{Watts1999} |D.~J. Watts|. ``Networks, dynamics, and the
  small-world phenomenon,'' !American Journal of Sociology\/!
  \<13>(2):493--527 (1999).

\bibitem{Weiss1999} |Y.~Weiss|. ``Segmentation using eigenvectors: A
  unifying view,'' in !Proc. of the Intl. Conf. Computer Vision\/!
  \<2>:975 (1999).

\bibitem{White2005} |S.~White| and |P.~Smyth|. ``A spectral clustering
  approach to finding communities in graphs,'' in !Proc. 5th SIAM
  Intl. Conf. Data Mining!, eds. |H.~Kargupta|, |J.~Srivastava|,
  |C.~Kamath|, and |A.~Goodman|. Philadelphia PA, pp.~76--84 (2005.)

\bibitem{Wolkowicz1999} |H.~Wolkowicz| and |Q.~Zhao|. ``Semidefinite
  programming relaxations for the graph partitioning problem,''
  !Discrete Applied Mathematics\/! \<96--97>:461--479 (1999).

\bibitem{Xu2008} |J. Xu| and |H. Chen|. ``The topology of dark
  networks,'' !Comm. ACM\/! \<51>(10):58--65 (2008).

\bibitem{Yang2007} |L.~Yang|. ``Data Embedding Research,'' Western
  Michigan University. Accessed 3~September 2012
  $\langle$\href{http://www.cs.wmich.edu/~yang/research/dembed/}{http://www.cs.wmich.edu/\discretionary{}{}{}\textasciitilde
    yang/research/dembed/}$\rangle$.

\bibitem{Yee2012} |M.~J. Yee|, |S.~Philips|, |G.~R. Condon|,
  |P.~B. Jones|, |E.~K. Kao|, |S.~T. Smith|, |C.~C. Anderson|, and
  |F.~R. Waugh|.  ``Network discovery with multi-source intelligence,
  surveillance, and reconnaissance,'' !Lincoln Laboratory~J.!,
  to~appear.

\end{thebibliography}

\endgroup 

\end{document}